\documentclass[fleqn,10pt]{wlscirep}
\usepackage[utf8]{inputenc}
\usepackage[T1]{fontenc}
\usepackage{multirow}
\usepackage{caption}
\usepackage{subcaption}
\usepackage{graphicx}
\usepackage{array}
\usepackage{float}
\title{The Intelligent ICU: Using Artificial Intelligence Technology for Autonomous Patient Monitoring}

\author[1]{Anis Davoudi}
\author[2]{Kumar Rohit Malhotra}
\author[2]{Benjamin Shickel}
\author[1]{Scott Siegel}
\author[3,4]{Seth Williams}
\author[3,4]{Matthew Ruppert}
\author[3,4]{Emel Bihorac}
\author[3,4]{Tezcan Ozrazgat-Baslanti}
\author[5]{Patrick J. Tighe}
\author[3,4,+]{Azra Bihorac}
\author[1,2,4,+,*]{Parisa Rashidi}

\affil[1]{Biomedical Engineering, University of Florida, Gainesville, 32611, USA}
\affil[2]{Computer and Information Science and Engineering, University of Florida, Gainesville, 32611, USA}
\affil[3]{Division of Nephrology, Hypertension and Renal Transplantation, Department of Medicine, College of Medicine, University of Florida, Gainesville, FL, USA}
\affil[4]{Precision and Intelligent Systems in Medicine (PrismaP), University of Florida, Gainesville, FL, USA}
\affil[5]{Department of Anesthesiology, College of Medicine, University of Florida, Gainesville, FL, USA}

\affil[*]{parisa.rashidi@bme.ufl.edu}

\affil[+]{these authors contributed equally to this work}


\begin{abstract}
Currently, many critical care indices are not captured automatically at a granular level, rather are repetitively assessed by overburdened nurses. In this pilot study, we examined the feasibility of using pervasive sensing technology and artificial intelligence for autonomous and granular monitoring in the Intensive Care Unit (ICU). As an exemplary prevalent condition, we characterized delirious patients and their environment. We used wearable sensors, light and sound sensors, and a camera to collect data on patients and their environment. We analyzed collected data to detect and recognize patient’s face, their postures, facial action units and expressions, head pose variation, extremity movements, sound pressure levels, light intensity level, and visitation frequency. We found that facial expressions, functional status entailing extremity movement and postures, and environmental factors including the visitation frequency, light and sound pressure levels at night were significantly different between the delirious and non-delirious patients. Our results showed that granular and autonomous monitoring of critically ill patients and their environment is feasible using a noninvasive system, and we demonstrated its potential for characterizing critical care patients and environmental factors.  
\end{abstract}
\begin{document}

\flushbottom
\maketitle
\thispagestyle{empty}

\section*{Introduction}
\begin{figure}[t]
\centering
\begin{subfigure}[t]{\textwidth}
\includegraphics[scale=0.25]{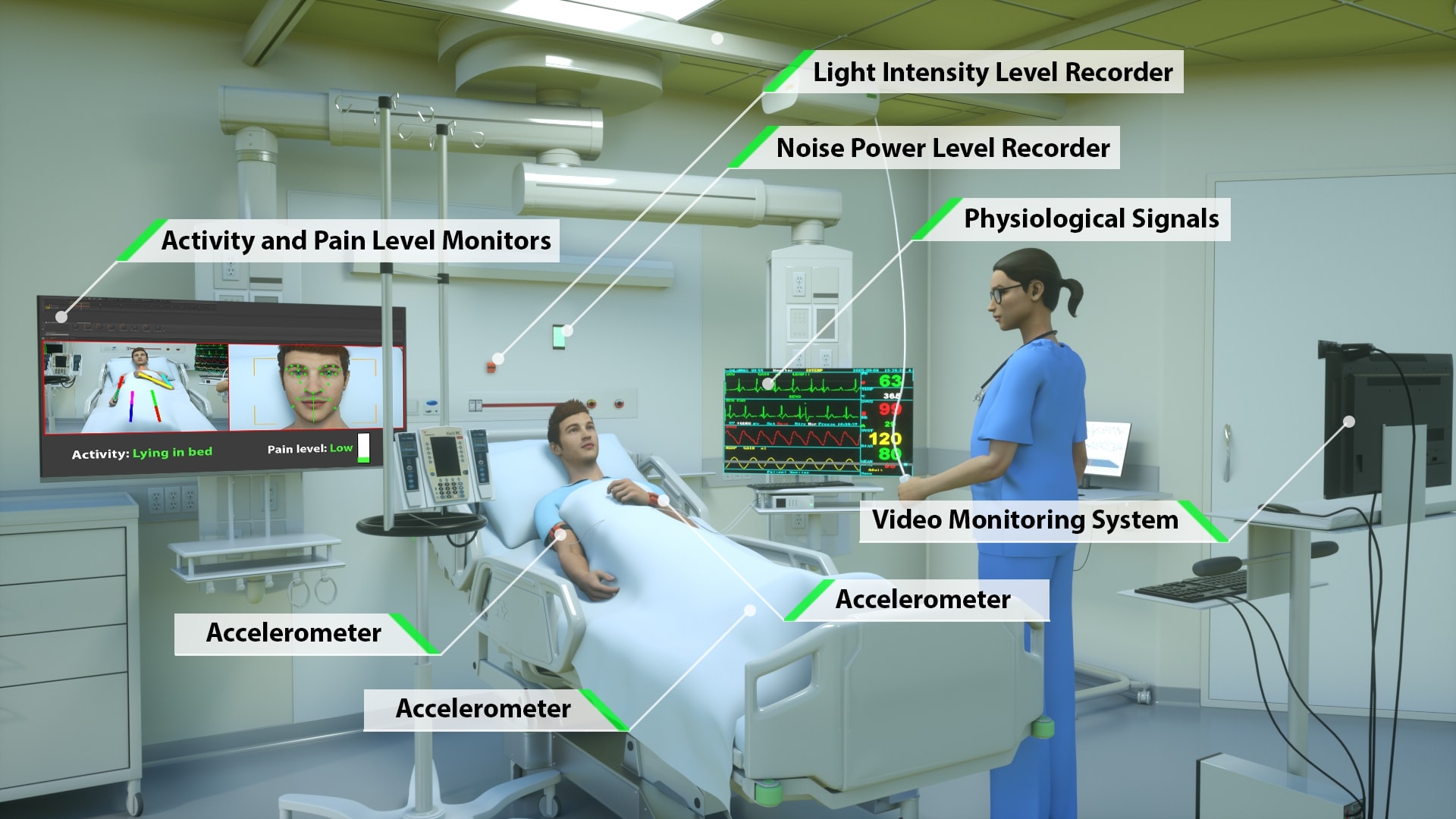}
\caption{}\label{fig:Figure_1a}
\end{subfigure}
\vspace{1cm}

\begin{subfigure}[t]{\textwidth}
\centering
\includegraphics[scale=0.15]{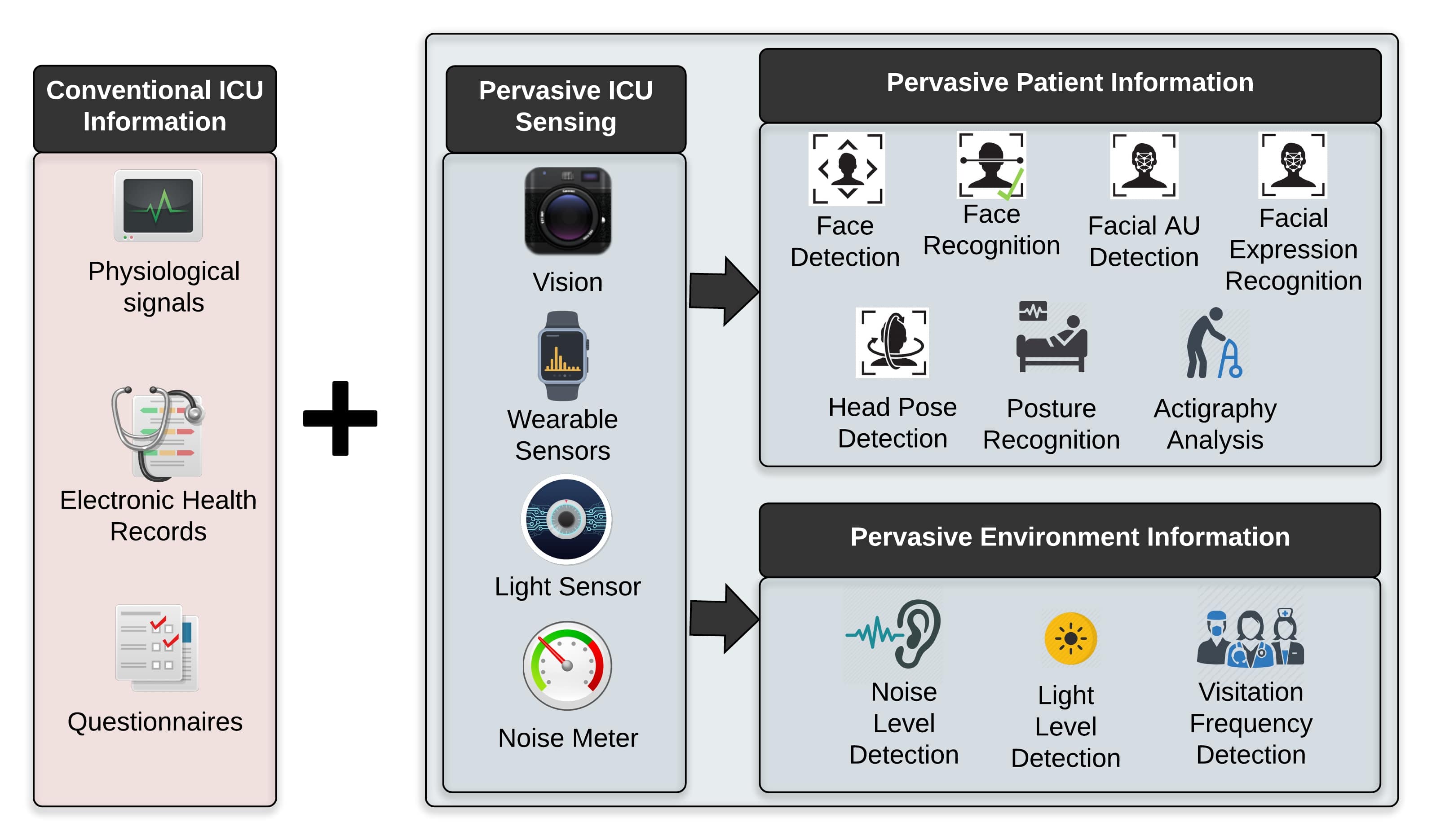}
\caption{}\label{fig:Figure_1b}
\end{subfigure}
\caption{(a) Intelligent ICU uses pervasive sensing for collecting data on patients and their environment. The system includes wearable accelerometer sensors, video monitoring system, light sensor, and a sound sensor. (b) The Intelligent ICU information complements conventional ICU information. Pervasive information is provided by performing face detection, face recognition, facial action unit detection, head pose detection, facial expression recognition, posture recognition, extremity movement analysis, sound pressure level, detection, light level detection, and visitation frequency detection.}
\end{figure}

Every year, more than 5.7 million adults are admitted to intensive care units (ICU) in the United States, costing the health care system more than 67 billion dollars per year \cite{RN506}. A wealth of information is recorded on each patient in the ICU, including high-resolution physiological signals, various laboratory tests, and detailed medical history in electronic health records (EHR) \cite{RN605}. Nonetheless, important aspects of patient care are not yet captured in an autonomous manner. For example, environmental factors that contribute to sleep disruption and ICU delirium \cite{RN803}, such as loud background noise, intense room light, and excessive rest-time visits, are not currently measured. Other aspects of patients’ well-being, including patient’s facial expressions of pain and various emotional states, mobility and functional status \cite{RN722,RN723} are not captured in a continuous and granular manner, and require self-reporting or repetitive observations by ICU nurses \cite{RN544,RN545}. It has been shown that self-report and manual observations can suffer from subjectivity, poor recall, limited number of administrations per day, and high staff workload. This lack of granular and continuous monitoring can prevent timely intervention strategies \cite{RN489, RN481, RN497,RN498,RN500,RN499}. 
With recent advancements in artificial intelligence (AI) and sensing, many researchers are exploring complex autonomous systems in real-world settings \cite{RNcar}.In ICU settings, doctors are required to make life-saving decisions while dealing with high level of uncertainty under strict time constraints to synthesize high-volume of complex physiologic and clinical data. The assessment of patients’ response to therapy and acute illness, on the other hand, is mainly based on repetitive nursing assessments, thus limited in frequency and granularity. AI technology could assist not only in administering repetitive patient assessments in real-time, but also in integrating and interpreting these data sources with EHR data, thus potentially enabling more timely and targeted interventions \cite{RN503, RN512}. AI in the critical care setting could reduce nurses’ workload to allow them to spend time on more critical tasks, and could also augment human decision-making by offering low-cost and high capacity intelligent data processing.
In this study, we examined how pervasive sensing technology and AI can be used for monitoring patients and their environment in the ICU. We utilized three wearable accelerometer sensors, a light sensor, a sound sensor, and a high-resolution camera to capture data on patients and their environment in the ICU (Figure \ref{fig:Figure_1a}\ref{fig:Figure_1b}). We used computer vision and deep learning techniques to recognize patient’s face, posture, facial action units, facial expressions, and head pose from video data. We also analyzed video data to find visitation frequency by detecting the number of visitors or medical staff in the room. To complement vision information for activity recognition, we analyzed data from wearable accelerometer sensors worn on the wrist, ankle, and arm. Additionally, we captured the room’s sound pressure levels and light intensity levels to examine their effect on patients’ sleep quality, assessed by the Freedman Sleep Questionnaire \cite{RN482}. For recruited patients, we retrieved all available clinical and physiological information from EHR (Table \ref{tab:Table1}). For a pilot study, we prospectively recruited 22 critically ill patients with and without ICU delirium to determine whether the Intelligent ICU system can be used to characterize the difference between their functional status, pain and environmental exposure. The Confusion Assessment Model-Intensive Care Unit (CAM-ICU) [19] was administered daily as the gold standard for detecting delirium.

\section*{Pilot Study Results}

We recruited 22 patients in the surgical ICU at the quaternary academic University of Florida Health Hospital (Table \ref{tab:Table1}, Figure \ref{fig:Figure_1_Supplement}) and five were excluded by the study’s exclusionary criteria. Nine patients (53\%) were diagnosed with ICU delirium by the daily CAM-ICU assessment for at least one day during the enrollment period. Delirious patients (defined as patients who were delirious throughout their enrollment period, number of patients=four) and non-delirious patients (defined as patients who were not delirious at any day during their enrollment period, number of patients=eight) did not significantly differ in baseline characteristics, except for the number of comorbidities (Table \ref{tab:Table1}). All delirious patients were identified as hyperactive subtype using the Delirium Motor Subtyping Scale (DMSS-4) \cite{RN838}. During the enrollment period, data were collected continuously for up to seven days from each patient. We collected 33,903,380 video frames visibly containing face, 16,123,925 video frames of patient posture, and 3,203,153 of patient facial expressions. We also collected 1,008 hours of accelerometer data, 768 hours of sound pressure level data, 456 hours of light intensity level data, and 1416 hours of physiological data. Occasionally, one or more sensors were removed at patient’s request, during bathing, or during clinical routines.  For training our deep learning models on ground truth labels, we annotated 65,000 video frames containing individual faces, and 75,697 patient posture video frames. All model training and testing were performed on an NVIDIA Titan X Pascal Graphical Processing Unit (GPU).

\subsection*{Face Detection}

To detect all individual faces in each video frame (including the patient, visitors, and clinical staff), we used the pretrained Joint Face Detection and Alignment using Multi-Task Cascaded Convolutional Network (MTCNN) \cite{RN369}. Face detection was evaluated on 65,000 annotated frames containing at least one individual face, resulting in a Mean Average Precision (mAP) value of 0.94.  
\subsection*{Patient Face Recognition}
To recognize patient face among detected faces, we implemented the FaceNet algorithm \cite{RN483} as an Inception-ResNet v1 model \cite{RN589}. The algorithm achieved an overall mAP of 0.80 and had slightly higher mAP value of 0.82 among non-delirious patients compared to delirious patients mAP of 0.75.

\subsection*{Patient's Facial Action Unit Detection}
We detected 15 facial action units (AUs) from 3,203,153 video frames using the pretrained OpenFace deep neural network  \cite{RN484}. The 15 AUs included six binary AUs (0 = absent, 1 = present), and 12 intensity-coding AUs (0 = trace, 5 = Maximum value), with three AUs reported as both binary and intensity (Table \ref{tab:Table 1_s}). SSuccessful detection was defined as the toolbox being able to detect the face and its facial AUs. Successful detection was achieved for 2,246,288 out of 3,203,153 video frames (\%70.1). The 15 detected AUs were compared between the delirious and non-delirious patients (Figure \ref{fig:Figure_2a}\ref{fig:Figure_2b}). All AUs were shown to be significantly different between the two groups (p-value<0.01).     

\subsection*{Facial Expression Recognition}

\begin{figure}
    \centering
    \begin{subfigure}[t]{\textwidth}
        \centering
        \includegraphics[scale=0.19]{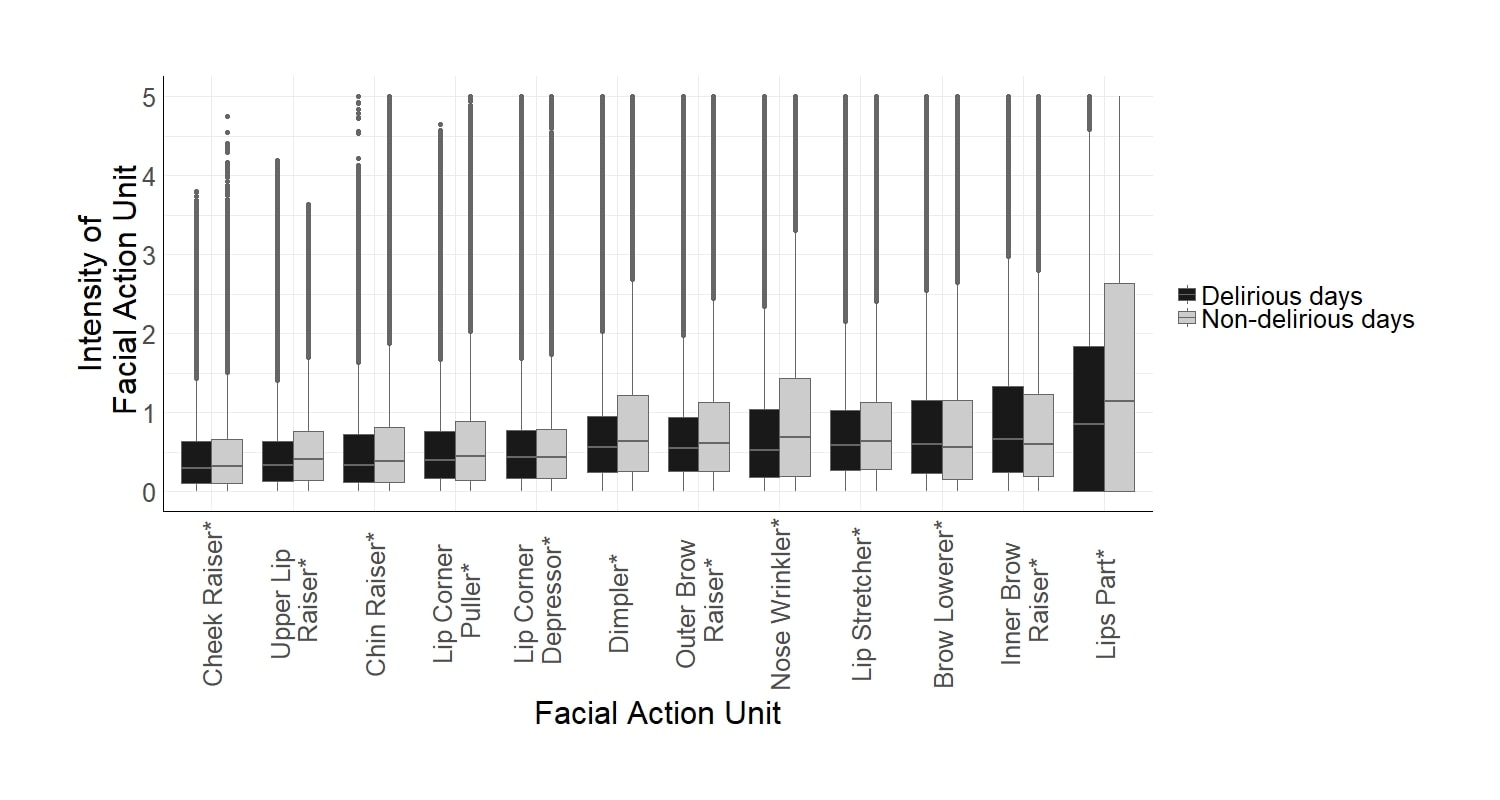} 
        \caption{} \label{fig:Figure_2a}
    \end{subfigure}
\hspace{0mm}
    \begin{subfigure}[t]{\textwidth} 
        \centering
        \includegraphics[scale=0.36]{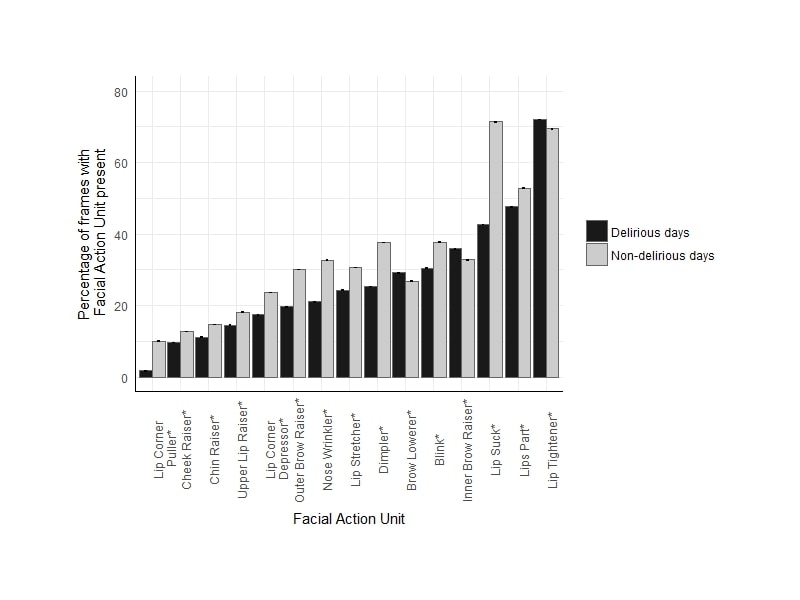} 
        \caption{} \label{fig:Figure_2b}
    \end{subfigure}
\hspace{0mm}
    \begin{subfigure}[t]{\textwidth}
    \centering
        \includegraphics[scale=0.19]{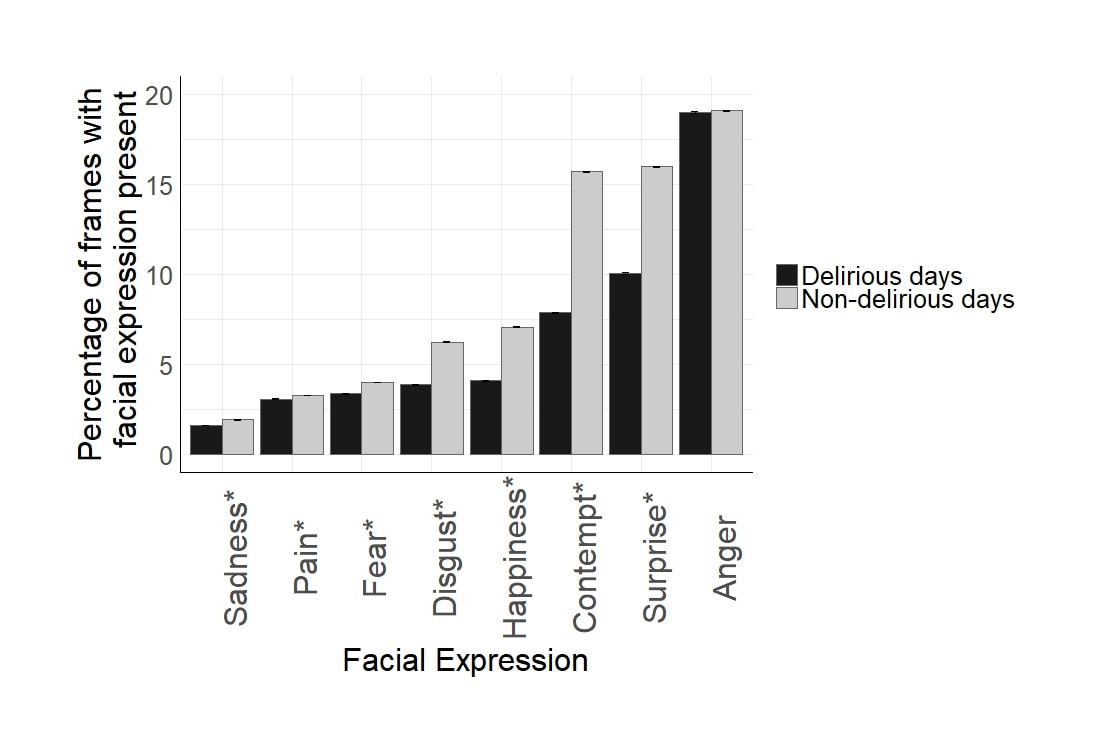} 
        \caption{} \label{fig:Figure_2c}
    \end{subfigure}
    \caption{a) Distribution of intensity-coding facial Action Units (AUs) among delirious and non-delirious patients shown as boxplots where middle line represents median and lower and upper end lines represents 25th and 75th percentiles, respectively. Facial AUs are coded between 0 (absence of the facial AU) to 5 (maximum intensity of facial AU); b) Percentage of frames with each binary-coding facial AU present among delirious and non-delirious patients during their enrollment period. Binary-coding facial AUs are coded either 0 (absent) or 1 (present). This bar plot shows how often a certain action unit is observed in all recorded video frames as percentage and standard error bars.; c) Percentage of frames with each facial expression present among the delirious and non-delirious patients, calculated based on constituent AUs (Table 2, Supplement). This bar plot shows how often a certain expression is observed in all recorded video frames as percentage and standard error bars. In a-c, * shows statistically significant difference between delirious and non-delirious groups (p-value<0.001).}
\end{figure}

We used the Facial Action Coding System (FACS) to identify common facial expressions from their constituent AUs (Table \ref{tab:Table 2_s}) \cite{RN602,RN603}. Eight common expressions were considered, including pain, happiness, sadness, surprise, anger, fear, disgust, and contempt. The occurrence rate of facial expressions was compared between the delirious and non-delirious patients (Figure \ref{fig:Figure_2c}). We were able to show that distributions of several facial AUs are different among the delirious and non-delirious groups. The differences in the distribution of such AUs point to the differences in affections of delirious and non-delirious patients. For instance, the presence of brow lowerer AU signals a negative valence \cite{RN815} and is stronger among the delirious patients than non-delirious patients (Figure \ref{fig:Figure_2a}\ref{fig:Figure_2b}). Facial expressions patterns can also potentially be used in predicting deterioration risks in patients  \cite{RN877}. Delirious patients had suppressed expression for seven out of eight emotions. All facial expressions except for anger had significantly different distribution among the delirious and non-delirious patients (p-value<0.001).

\subsection*{Head Pose Detection}

\begin{figure}
    \centering
    \setlength{\belowcaptionskip}{-4pt}
    \begin{subfigure}[t]{\textwidth}
        \centering
        \includegraphics[scale=0.3]{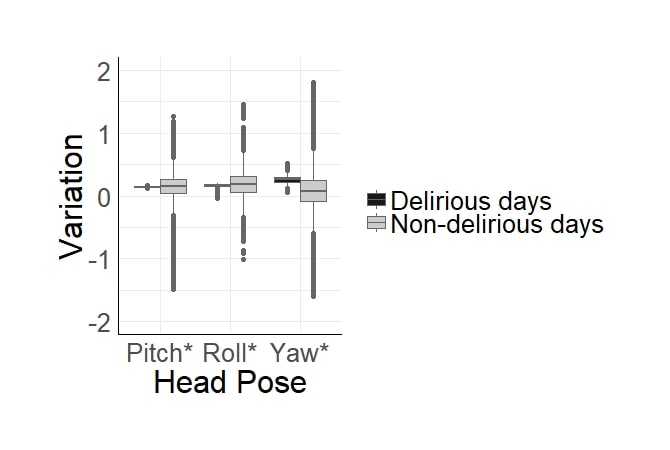} 
        \caption{} \label{fig:Figure_3a}
    \end{subfigure}
   \vspace{0mm}
   \setlength{\belowcaptionskip}{-4pt}
     \begin{subfigure}[t]{\textwidth}
        \centering
        \includegraphics[scale=0.28]{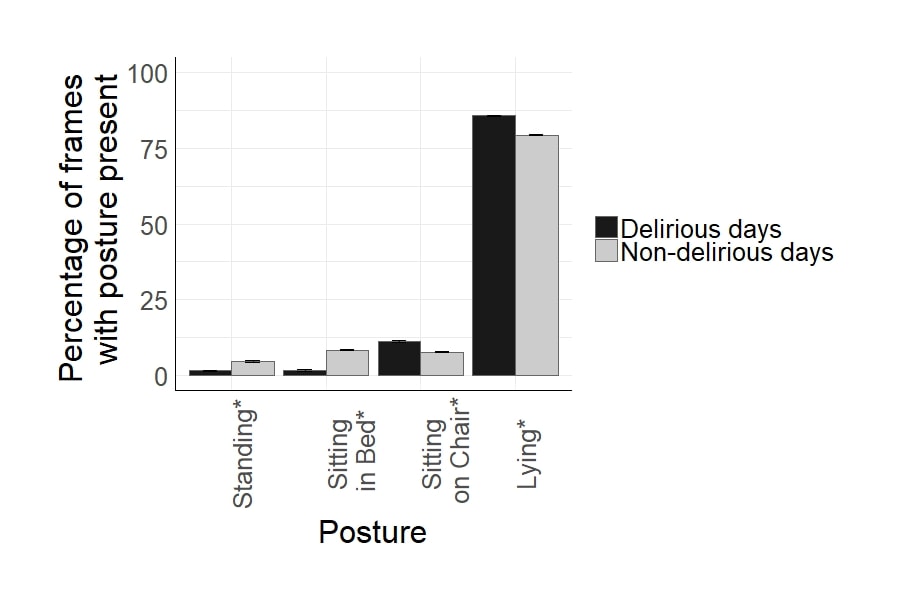} 
        \caption{} \label{fig:Figure_3b}
    \end{subfigure}
    \vspace{0mm}
    \setlength{\belowcaptionskip}{-4pt}
	\begin{subfigure}[t]{\textwidth}
    \centering
        \includegraphics[scale=0.58]{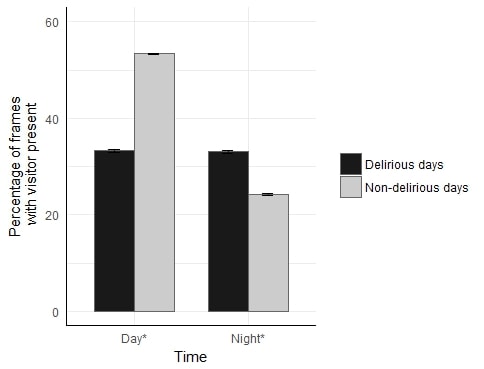} 
        \caption{} \label{fig:Figure_3c}
    \end{subfigure}
    \caption{a) Distribution of head poses among delirious and non-delirious patients during their enrollment days shown as boxplots. Pitch, yaw, and roll describe the orientation of the head in its three degrees of freedom. Pitch is the rotation around the right-left axis, up and down, as shaking the head “Yes”. Roll is rotation around the inferior-superior axis, as shaking the head “No”. Yaw is rotation around the anterior-posterior axis, side to side, like shaking the head “Maybe”; b) Percentage of the frames spent in each posture among delirious and non-delirious patients shown along with standard error bars; c) Percentage of frames with visitors present in the room (disruption) for delirious and non-delirious patients during the day (7AM-7PM) and during the night (7PM-7AM) shown along with standard error bars. In a-c, * shows statistically significant difference between delirious and non-delirious groups (p-value<0.001).}
\end{figure}

We detected three head poses including yaw, pitch, and roll, using the pretrained OpenFace deep neural network tool \cite{RN484}. The head rotation in radians around the Cartesian axes was compared between the delirious and non-delirious patients, with the left-handed positive sign convention, and the camera considered as the origin. Delirious patients exhibited significantly less variation in roll head pose (rotation in-plane movement), in pitch head pose (up and down movement), and in yaw head pose (side to side movement) compared to the non-delirious patients (Figure \ref{fig:Figure_3a}). Extended range of head poses in non-delirious patients compared to delirious patients (Figure \ref{fig:Figure_3a}) might be the result of more communication and interaction with the surrounding environment.

\subsection*{Posture Recognition}

To recognize patient posture, we used a multi-person pose estimation model \cite{RN493} to localize anatomical key-points of joints and limbs. Then we used the lengths of body limbs and their relative angles as features for recognizing lying in bed, standing, sitting on bed, and sitting in chair. We obtained an F1 score of accuracy of 0.94 for posture recognition. The highest misclassification rate (11.3\%) was obtained for sitting on chair (misclassified as standing). The individual classification accuracy of recognizing postures was: lying = 94.5\%, sitting on chair = 92.9\%, and standing = 83.8\% (Table \ref{tab:Table 3_s}). Delirious patients spent significantly more time lying in the bed and sitting on chair compared to and non-delirious patients (p-value<0.05 for all four postures, Figure \ref{fig:Figure_3b}).

\subsection*{Extremity Movement Analysis}

We  analyzed the data from three accelerometer sensors worn on patient’s wrist, ankle, and arm, and compared the results between delirious and non-delirious patients. For the purpose of feature calculation, we consider daytime from 7 AM to 7 PM, and nighttime from 7 PM to 7 AM, based on nursing shift transitions. Figures  \ref{fig:Figure_4} show the smoothed accelerometer signal averaged over all delirious and all non-delirious patients. We also derived 15 features per each accelerometer (Table \ref{tab:Table2}), resulting in 45 total features for the wrist, ankle, and arm sensors. We compared the extracted features in delirious and non-delirious patients for the wrist-worn sensor, arm-worn sensor, and ankle-worn sensor  (Table \ref{tab:Table2}, \ref{tab:Table 4_s}\ref{tab:Table 5_s}). 
Delirious patients had higher movement activity for wrist and lower extremity, and lower movement activity for the upper extremity during the entire 24-hours cycle, daytime (7 AM-7 PM), and nighttime (7 PM-7 AM). The 10-hour window with maximum activity intensity showed different levels of activity between the two patient groups. However, activity in the 5-hour window with the lowest activity intensity was not significantly different, possibly due to low activity levels in ICU in general. The number of immobile moments during the day and during the night were also different between the two groups, with less number of immobile moments detected for the delirious patients, hinting at their restlessness and lower sleep quality. The extremity movement features did not show significant difference for arm and ankle. This might stem from the overall limited body movements of all ICU patients.
\subsection*{Visitation Frequency}

The pose estimation model was also used on the video data to identify the number of individuals present in the room at any given time, including visitors and clinical staff. Delirious patients on average had fewer visitor disruptions during the day, but more disruptions during the night (Figure \ref{fig:Figure_3c}). 

\subsection*{Room Sound Pressure Levels and Light Intensity}

\begin{figure}[ht]
\centering
\includegraphics[scale=0.19]{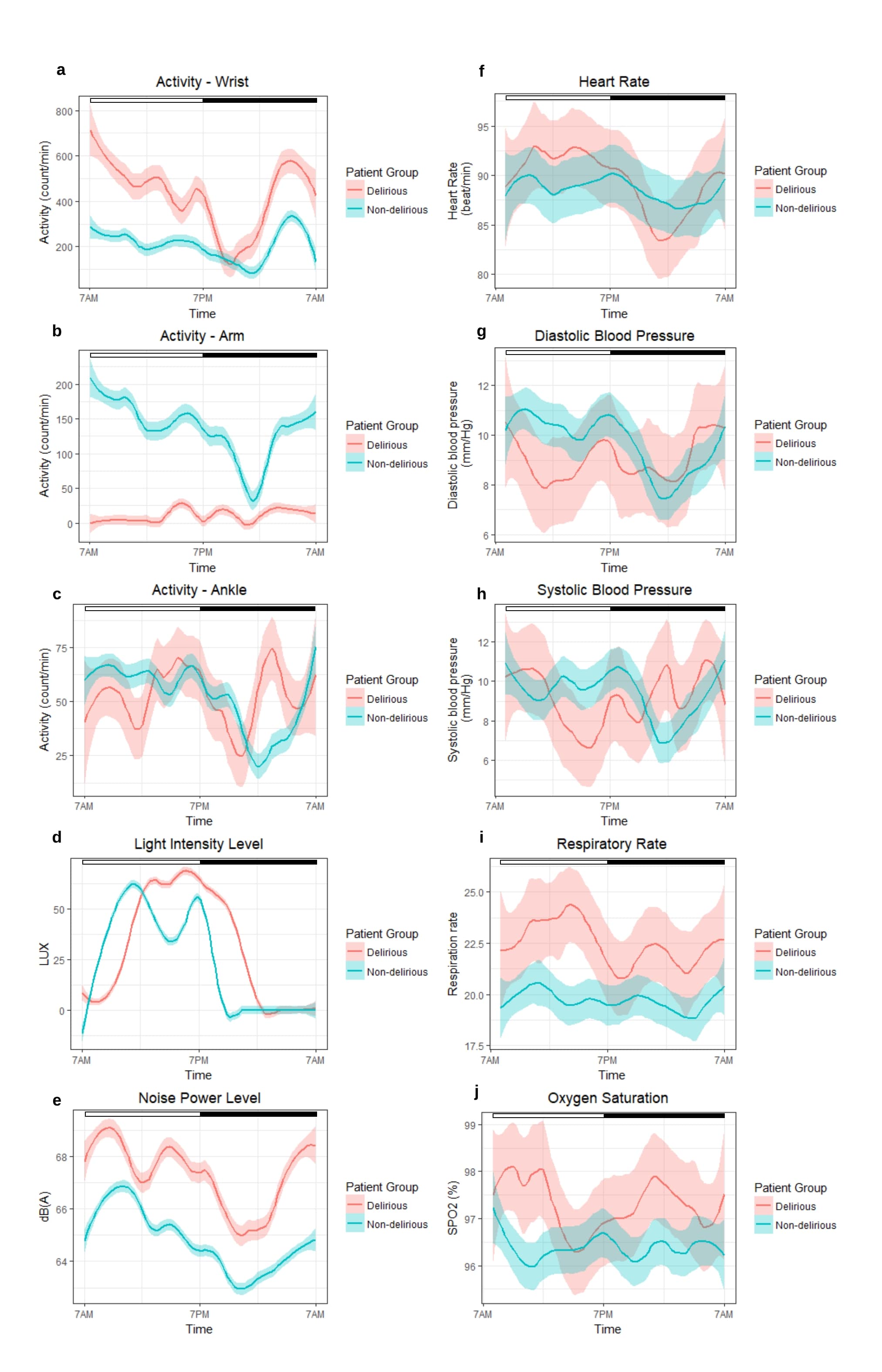}
\caption{Delirious and non-delirious group comparisons for a-e) sensor data and f-j) physiological data. Sensor data included accelerometer data recorded on the wrist, arm, and ankle, as well as light intensity level recorded using an actigraph capable of recording light intensity level and the sound pressure level using an iPod on the wall behind the patient’s bed physiological data included heart rate, systolic blood pressure and diastolic blood pressure, respiration rate, and oxygen saturation. Physiological data were collected with a resolution of approximately once per hour as part of the patient’s care. The graphs show the smoothed average value per group, with the transparent band around each average line showing the 95\% confidence interval.}
\label{fig:Figure_4}
\end{figure}

The sound pressure levels for delirious patients’ rooms during the night were on average higher than the sound pressure levels of non-delirious patients’ rooms (Figure \ref{fig:Figure_4}). Average nighttime sound pressure levels were significantly different between the delirious and non-delirious patients (p-value<0.05). Delirious patients on average experienced higher light intensity during the evening hours, as can be seen in Figure \ref{fig:Figure_4}. Average nighttime light intensity levels were significantly different between the delirious and non-delirious patients (p-value<0.05).

\subsection*{Sleep Characteristics}

We examined the Freedman Sleep Questionnaire \cite{RN482} responses for the delirious and non-delirious patients to compare their sleep patterns. While the median of the overall quality of sleep in the ICU and effect of acoustic disruptions and visitations during the night were different among the delirious and non-delirious groups, these differences were not statistically significant. However, delirious patients reported a lower overall ability to fall asleep compared to non-delirious patients, and they were more likely to find the lighting to be disruptive during the night (p-value= 0.01, p-value=0.04, respectively, Figure \ref{fig:Figure_2__Supplement}). 

\subsection*{Physiological and EHR data}

Patients’ demographic and primary diagnosis were not significantly different between the delirious and non-delirious patients  (Table \ref{tab:Table1}). Delirious patients on average had higher average heart rate, oxygen saturation, and respiration rate, a sign of potential respiratory distress and agitation. Systolic and diastolic blood pressure of the delirious patients were lower than non-delirious patients during the evenings (Figure \ref{fig:Figure_4}). All delirious patients received continuous enteral feeding orders and were fed throughout the nighttime while 50\% of non-delirious patients had enteral feeding order during their enrollment days.  

\section*{Discussion}

In this study, we showed the feasibility of pervasive monitoring of patients in the ICU. This is the first study to develop an autonomous system for patient monitoring in the ICU. We performed face detection, face recognition, facial action unit detection, head pose detection, facial expression recognition, posture recognition,  extremity movement analysis, sound pressure level detection, light level detection, and visitation frequency detection, in the ICU. As an example, we evaluated our system for characterization of patient and ambient factors relevant to delirium syndrome \cite{RN803,RN801, RN802}. Such a system can be potentially used for detecting activity and facial expression patterns in patients. It also can be used to quantify modifiable environmental factors such as noise and light in real time. This system can be built with an estimated cost of < \$300 per ICU room, a relatively low cost compared to daily ICU costs of thousands of dollars per patient. It should be noted that after proper cleaning procedures, the same devices can be also reused for other patients, further reducing the amortized cost per patient in the long term.

To the best of our knowledge, this is the first study to continuously assess critically ill patients’ affect and emotions using AI. The AI introduces the ability to use the combination of these features for autonomous detection of delirium in real time and would offer a paradigm shift in diagnosis and monitoring of mood and behavior in the hospital setting. 

Our system also uniquely offers autonomous detection of patients’ activity patterns by applying deep learning on sensor data obtained from video and accelerometer. This previously unattainable information can optimize patients’ care by providing more comprehensive data on patients’ status through accurate and granular quantification of patients’ movement. While there is previous work that has used video recordings in the ICU to detect patient’s status \cite{RN547}, they were not able to measure the intensity of patients’ physical activity. The combined knowledge of patients’ functional status through video data and their physical activity intensity through movement analysis methods can help health practitioners to better decide on rehabilitation and assisted mobility needs. 

Our collected data hint at several interesting observations, including more significant disruption of the circadian rhythm of physical activity in delirious patients, as confirmed by other studies \cite{RN591,RN781,RN783}. To concur with existing practice in literature, we used standardized tools accepted as gold standard including Friedman’s sleep quality questionnaires and CAM-ICU assessment. In contrast to previous studies, we collected and analyzed information at a granularity and with an accuracy that is impossible to obtain using conventional methods such as questionnaires. Circadian rhythm, which is important for many health regulatory processes in the body, is generally severely disrupted in ICU patients \cite{RN591,RN781,RN783}. Delirious patients reported lower overall ability to fall asleep in the ICU. Delirious patients also reported a higher degree of disruption from lighting compared to non-delirious patients. Delirious patients’ noise perception was not statistically significantly different from the non-delirious group, even though the average sound pressure level from the delirious patients’ rooms was higher -almost equivalent to street traffic noise during the sleep time. This could be possibly due to affected hearing and vision perception of delirious patients. As shown in other studies on noise and light exposures in the ICU, both delirious and non-delirious patients experienced sound pressure levels that were well above the recommended guidelines of World Health Organization (WHO) (Figure 4): 35 dB, with a maximum of 40 dB overnight \cite{RN872, RN873}. Continuous measurement of noise and light exposures in patients in the ICU allow for implementing real-time interventions to improve patients’ sleep hygiene.

Using our vision techniques, we were also able to observe differences in visitor disruptions among delirious and non-delirious patients. More disruptions were observed during the day for non-delirious patients. This may both contribute to and stem from delirium since interactions with others can have reorientation effects and reduce the risk of delirium. At the same time, since delirious patients do not engage in interactions with others, they might have shorter visitations from both family and clinical staff. This was further corroborated by the smaller variation in head pose variation among delirious patients compared to non-delirious patients. More disruptions during the day for non-delirious patients possibly points to their capability to have more interactions with others, including family caregivers and visitors. ICU environment instills a sense of loneliness in the patients, reported by many delirious patients \cite{RN596, RN597, RN598, RN541}. Visitations during the day may contribute to preventing delirium, because of conversations’ reorientation effects and the sense of relief that seeing recognizable faces invokes. On the other hand, isolation precautions have been reported to up to double the rate of delirium \cite{RN594, RN595, RN600}. Delirious patients also had more disruptions during the night, possibly due to their more severe condition and frequent visits from providers, which could aggravate their sleep disruptions.

Using our extremity movement data, we showed that patients’ activity in the wrist was significantly different between delirious and non-delirious patients, but this was not the case for the arm and ankle. It should be noted that the ideal location for wearing the actigraphy device is not necessarily the same for different clinical conditions \cite{RN812, RN813, RN814}. This insight can potentially be used to reduce the number of required on-body sensors for monitoring specific conditions such as delirium. To the best of our knowledge, this is the first study that examines the wear location for actigraphy devices in characterizing activity patterns in delirium patients. The wrist-based extracted features can be potentially used in delirium detection, as well as in tracking the efficacy of mobility and rehabilitation interventions.

While our system shows great promise for future ICU patient monitoring applications, there are several limitations that need to be considered. As a pilot study to assess the feasibility of the system, we recruited a small number of patients. This could possibly affect some of the results and their variability, once examined in a larger cohort. For example, we did not detect any significant differences between movement features derived from ankle and arm sensors, but larger sample size and more diverse features might show higher discriminative power. Using a larger sample size would also allow us to better customize our deep learning vision models to the ICU environment. 

Another limitation of the study is that we did not consider the medications that patients were receiving during their stay in the ICU. Sedatives, analgesics, and anticholinergic medications may potentially affect patients’ sleep, pain, activity patterns, and delirium. To correctly account for these factors, the drugs’ dosages and half-times need to be extracted and their continuous effect on each individual patient needs to be considered. 

Implementing our system in a real ICU also poses many challenges, resulting in several limitations in our study. One major issue is the crowded scene in the ICU room, resulting in patient face occlusion and inability to detect expressions at all times. One potential solution is deploying multiple cameras with different view angles. Patient’s face and body might be also obstructed by ventilation devices, bandage, or simply blocked from view by blankets. Another issue arises from the multitude of medical devices on the patient, making it difficult to use wearable sensors on all patients. A related issue occurring for the wearable sensors is that they might get lost, or the clinical staff might not place them in the right location after taking them off for a bath. 

Another challenge was inaccurate environment data collection, stemming from the physical properties of light waves. While we made every effort to place the light sensor near the patient’s head to record the same amount of light exposure that the patient is experiencing, individuals or medical equipment might block the light sensor. Again, deploying multiple light sensors, as well as developing vision-based light analysis modules can alleviate this problem.

Privacy can be also a major concern for any system that uses video monitoring. To be able to develop and validate our system, we needed to annotate the video frames to establish ground truth for every event -face recognition, posture detection, and disruptions. However, a future operational version can rely on real-time and online vision analysis without storing any video data. This approach could also reduce the need for extensive storage requirements.

In summary, as a proof of concept, we were able to demonstrate the feasibility of pervasive monitoring of patients in the ICU. We expect future similar systems can assist in administering repetitive patient assessments in real-time, thus potentially enabling more accurate prediction and detection of negative events, and more timely interventions, reducing nursing workload, and opening new avenues for characterizing critical care conditions on a much more granular level. 

\section*{Materials and Methods}

This study was carried out at surgical ICUs in the University of Florida Health Hospital, Gainesville, Florida. It was approved by University of Florida Institutional Review Board by IRB 201400546, and all methods were performed in accordance with the relevant guidelines and regulations. Written informed consent was obtained from patients and/or their surrogates before enrollment in the study. All adult patients 18 years or older who were anticipated to stay longer than 48 hours in ICU and were able to wear an ActiGraph were eligible for enrollment. For each patient, camera and accelerometer data, along with surveys data were collected for up to one week, or until discharge or transfer from the ICU, whichever occurred first. EHR data were collected for the duration of patients’ hospitalization. All history data for diagnoses and procedures were also retrieved for analysis.

Among consenting patients, two patients withdrew before data collection started, one was excluded because of being transferred from the ICU before data collection commenced, two patients stayed for less than a complete day in the ICU. Five patients were not considered in the delirium analysis because they had both delirium and non-delirium days. 

For characterizing delirium in terms of activity, facial expressions, environmental light and sound pressure levels, and disruption, delirious patients were defined as those who were delirious throughout their enrollment period, as detected by the CAM-ICU test. DMSS-4 and Memorial Delirium Assessment Scale (MDAS) \cite{RN875} tests were used to detect the motoric subtype and severity of delirium, respectively. To better observe the differences between the delirious and non-delirious patients, we only compared the data between patients who were delirious throughout their enrollment period and those who were non-delirious throughout their enrollment period.  

\subsection*{Data Acquisition}
The pervasive sensing system for data acquisition included (1) a high-resolution and wide-field-of-view camera, (2) three wearable accelerometer sensors, (3) light sensor, (4) microphone for capturing sound pressure levels, and (5) a secure local computer. A touchscreen user-friendly interface allowed nurses and caregivers to stop data collection at any time. Data were captured on a local secure computer throughout the patient enrollment period and transferred to a secure server for analysis upon patient discharge. 
\subsubsection*{Vision}
We captured video using a camera with a 90o diagonal field of view with 10X optical zoom for zooming on the patient face, and at 15 frames per seconds (fps) speed. The camera was placed against the wall, facing the patient. We took several privacy precautions with respect to video data, including posting clearly visible signs warning of “recording in session”. A sliding lens cover was used as a quick privacy alternative. No audio information beyond aggregate sound pressure levels was collected. A simple user interface also allowed the nurses and family caregivers to stop recording at any time, or to delete any scenes if needed.  

\subsubsection*{Wearable Sensors}
Wearable accelerometer sensors provide complementary information to vision information for activity recognition in the ICU. We used three Actigraph GT3X (GT3X) devices (ActiGraph, LLC. Pensacola, Florida) to record patients’ activity intensity throughout their enrollment period. We placed one GT3X device on the patient’s dominant wrist, one on the dominant arm, and another on the dominant ankle to be able to examine different lower and upper body movements. For patients with medical equipment on their dominant wrist, arm, or ankle, or those who did not wish to wear the device on any of these positions, we placed the GT3X device on the opposite side, or removed it altogether, if necessary. Nurses were instructed to remove the devices for bathing and medical procedures, if necessary, and to replace the devices afterward. The GT3X sensor weighs less than 19 grams and can record data for up to 25 days. It records activity intensity in form of activity counts \cite{RN887}. We recorded data at 100Hz sampling rate and used 1-min activity counts in our analysis.  
\subsubsection*{Sound and Light}
To capture the effect of environment disruptions on sleep quality, we recorded light intensity and sound pressure levels in the room throughout the patient’s enrollment period. 
\subsubsection*{Physiological Signals and EHR Data}
We retrieved electronic health records EHR data using the University of Florida Health (UFH) Integrated Data Repository as Honest Broker. EHR data included physiological signals recorded in the ICU via bedside monitors, including heart rate, temperature, systolic and diastolic blood pressure, respiratory rate, and oxygen saturation. Additionally, we retrieved EHR information on demographics, admission information, comorbidities, severity scores, pain and CAM-ICU scores, laboratory results, medications, procedure and diagnosis codes, and enteral feeding status.
\subsubsection*{Questionnaires}
As the gold standard, we administered daily questionnaires to assess patients’ sleep quality during their enrollment. We used the Freedman sleep questionnaire. We also administered the CAM-ICU delirium assessment daily. For delirious patients, we administered MDSS-4 and MDAS tests daily to identify the subtype and severity of delirium. For patients who were asleep or unavailable because of clinical procedures, the questionnaires were administered as soon as they became available, before 3 pm. 

\subsection*{Analysis}
Once data was captured by our pervasive sensing system, it was analyzed to examine different aspects of patient status and environment. We used vision information for face detection, face recognition, facial action unit detection, head pose detection, facial expression recognition, posture recognition, and visitation frequency detection. We employ several deep neural network algorithms for analyzing video data and we use statistical analysis to analyze accelerometer sensor data, as well as to compare the environmental factors and activity intensity of the delirious and non-delirious patients. For numerical variables description, we used mean and standard deviation in case of approximately normal distribution. In case of variables with skewed distribution, we used median and interquartile range (25th, 75th). We used Python: 2.7, Tensorflow: 1.1.0, OpenCV: 3.2.0, Caffe: 1.0 for implementing vision algorithms. We used Scikit-learn: 0.18.1 and R: 3.4.1 for conventional machine learning and statistical analysis.

\subsubsection*{Face Detection}
As a first step, we detected all individuals present in the room. We used the Joint Face Detection and Alignment using Multi-Task Cascaded Convolutional Network (MTCNN) to detect individuals in each video frame. This framework employs a cascaded architecture with three stages of deep convolutional neural networks (CNN) to predict face and landmark locations in a coarse-to-fine manner (Figure \ref{fig:Figure_3__Supplement}). We evaluated the accuracy of the MTCNN models based on ground truth provided by expert annotator. A total of 65,000 video frames were annotated by delineating a bounding box surrounding each individual. 
\subsubsection*{Face Recognition}
After individual faces were detected, we performed face recognition to identify the patient in each video frame. This step is necessary since at any given moment several individuals can be present in the room, including the patient, nurses, physician, and visitors. To perform face recognition, we implemented the FaceNet algorithm, which consists of an Inception-ResNet V1 model. First, we extracted 7 seconds of still images at 15 fps containing the patient face, as training data. Training data was passed through the face detection pipeline. Thus, the input to FaceNet model is the set of aligned images obtained from MTCNN. The trained classifier for each patient was tested on 6,400 randomly selected images of the same patient, containing both patients and non-patients in the same frame. We evaluated the accuracy of the FaceNet model per each patient. If a patient was recognized with a probability of 0.9 or higher, it was reported as a positive recognition. Pipeline of the patient face recognition system is shown in Figure \ref{fig:Figure_5a}. 
\subsubsection*{Facial Action Unit and Expression Recognition}
For each video frame, facial Action Units (AUs) were obtained from the OpenFace toolbox and were used to detect eight common facial expressions: pain, happiness, sadness, surprise, anger, fear, disgust, and contempt (Table  \ref{tab:Table 1_s}). Facial expressions were only considered during daytime (7 AM-7 PM), when there is sufficient light in the room. Based on the FACS formulas (Table \ref{tab:Table1}), pain expression was identified as a combination of the following AUs: AU4 (brow lowerer), AU6 (cheek raiser), AU7 (lid tightener), AU9 (nose wrinkler), AU10 (upper lip raiser), and AU43 (eyes closed). Other facial expressions can be similarly identified. Once an expression was detected, we computed how often that expression $e_i$ was observed, denoted as the expression frequency $f_i$. The relative expression frequency $f_i$ is calculated as in equation (1). Here, $N_i$ refers to the number of frames where expression $e_i$ was observed, and \textit{N} refers to the total number of frames.
\begin{equation} f_i =\frac{N_i}{N}    \end{equation}

\subsubsection*{Posture Classification}
After detecting and recognizing patient’s face, we localized anatomical key-points of joints and limbs using the real-time multi-person 2D pose estimation \cite{RN493} with part affinity fields (Figure \ref{fig:Figure_5b}). This allowed us to recognize poselets, which describe a particular part of posture under a given viewpoint \cite{RN800}. The part affinity fields are 2D vector fields that contain information about the location and direction of limbs with respect to body joints. Our pose detection model consisted of two Fully Convolutional Neural networks (FCN) \cite{RN718} branches, where one branch detects the location of the joints, and the other branch detects the association of those body joints as limbs. Identified poselets were provided to a k-nearest neighbor (KNN) classifier to identify the full posture. To train the model using a balanced dataset, we augmented ICU patient data with scripted data (Appendix B). We considered four main posture classes to be recognized: lying in bed, standing, sitting on bed, and sitting on chair.  Several ICU functional and mobility scales are based on evaluating patients’ ability to perform these activities \cite{RN886, RN885}.
\subsubsection*{Extremity Movement}
We calculated several statistical features to summarize the accelerometer data obtained from the wrist, arm, and ankle. We used the LOESS non-parametric regression method for smoothing to show the smoothed average of activity intensity for each group over the course of the day. We compared fifteen features extracted from the accelerometer data of the delirious and non-delirious patients. Features used in this analysis include four groups of features to reflect different aspects of activity intensity patterns. First group of features includes mean and standard deviation (SD) of activity intensity for the whole day, during daytime, and during nighttime. Second group of features includes activity intensity of 10-hour window with highest sum of activity intensity (M10), time of start of M10 window, activity intensity of 5-hour window with lowest sum of activity intensity (L5), time of start of  L5 window, relative amplitude- which is (M10-L5)/(M10+L5). Third group of features consists of Root Mean Square of Sequential Differences (RMSSD), RMSSD/SD. The fourth group of features includes the number of immobile minutes during daytime and during nighttime. Immobile minutes are defined as minutes with activity intensity of zero. Mean of activity intensity reflects the amount of patient’s activity intensity, while standard deviation of activity intensity reflects the amount of change in patient’s movement. M10 and L5, and their corresponding start times are chosen to show what times the most amount of activity and the least amount of activity occur, and to detect whether they correspond to daytime and nighttime, respectively. RMSSD is used to detect immediate changes in the activity intensity, which can potentially point to unintentional activity, since patients in the ICU do not normally have fast actions for long periods of time, and RMSSD/SD normalizes these immediate changes by the overall changes in the data captured by standard deviation. Number of immobile minutes can be used to observe the percentage of the time patients were inactive during the day (undesirable) and during the night (desirable). 
\subsubsection*{Sound and Light}
We placed an iPod with a sound pressure level recording application and a GT3X with a light sensor on the wall behind the patient’s bed to record sound pressure levels and light intensity levels (Table \ref{tab:Table 6_s}). 

Sound pressure level collection was performed using the built-in microphone of the iPod. Sound waves can be described as a sequence of pressure changes in the time domain, and ears detect changes in the sound pressure \cite{RN842}. Sound pressure level (SPL) is a logarithmic measure of the effective pressure of a sound relative to a reference value and is measured in decibel (dB). Sound pressure is proportional to sound intensity and is defined as in equation (2).

\begin{equation} SPL =ln(\frac{p}{p_0}){N_p}=2{log_{10}}(\frac{p}{p_0})B=20{log_{10}}(\frac{p}{p_0})dB   \end{equation}

\bibliography{sample}

\section*{Acknowledgements}

This work is supported by NSF CAREER 1750192 (PR), NIH/NIGMS P50 GM111152 (AB, TOB), NIH/NIA P30 AG028740 (PR), and NIH/NIGMS RO1 GM-110240 (PR, AB). The Titan X Pascal partially used for this research was donated by the NVIDIA Corporation. 

\section*{Author contributions statement}

Study design: A.B. and P.R; Data collection: M.R., S.W., E.B., and A.D.; Data analysis: A.D., K.R.M., B.S., S.S., and T.O.; Data interpretation: A.D., K.R.M., and P.J.T.; Drafting the manuscript: A.D. and P.R.; Critical revision of the manuscript: A.B. and P.R. all authors approved the final version of the manuscript to be published.  

\section*{Additional information}
\textbf{Competing interests}
The authors declare no competing interests. 
\section*{Appendix A}
\subsection*{Face Detection}
To detect individual faces, we extracted seven seconds of still images at 15 fps as training data and used the Joint Face Detection and Alignment using Multi-Task Cascaded Convolutional Network (MTCNN). This framework employs a cascaded architecture with three stages of deep convolutional neural networks (CNN) to predict face and landmark locations in a coarse-to-fine manner. In the first stage, candidate windows possibly containing faces are produced using a fully convolutional network called Proposal Network (P-Net) (Figure \ref{fig:Figure_2__Supplement}) \cite{RN588}. Each candidate window has four coordinates – top left coordinates, height, and width. Ground truth bounding boxes have the same coordinate format as well. The objective function for bounding box regression performed on these candidate windows is the Euclidean loss between the corresponding coordinates of a candidate window and its nearest ground truth bounding box. The objective is to minimize this Euclidean loss, given for a sample $x_i$ as in equation (1).
\begin{equation}
L_i^{box}=||\hat{y}_i^{box}-y_i^{box}||_2^2
\end{equation}
Here,\( \hat{y}_i^{box}\) is the output regression coordinate obtained from the network and \(y_i^{box}\) is the ground-truth coordinate. After performing bounding box regression, the highly overlapping candidates are merged using non-maximum suppression (NMS) \cite{RN817}. NMS is performed by sorting the bounding boxes by their score, and greedily selecting the highest scoring boxes and removing the boxes that overlap with the already selected boxes more than a given threshold, 0.7 in the first stage. In the second stage, all candidates selected in the first stage are provided to another convolutional network, Refine Network (R-Net) (Figure \ref{fig:Figure_2__Supplement}). R-Net further rejects candidate windows not containing faces, performs bounding box regression, and merges the NMS candidates with a threshold of 0.7. Finally, the Output Network (O-Net) produces the final bounding box (Figure \ref{fig:Figure_2__Supplement}). MTCNN is trained for bounding box regression by posing its objective function as a regression problem. While extracting the candidate windows during testing, a window is selected on the basis of the threshold given for Intersection over Union (IoU) score, calculated as in equation (2).
\begin{equation}
IoU_i=\frac{A_i^o}{A_i^u}
\end{equation}
Here, \(A_i^o\) is the area of overlap between the \textit{ith} ground-truth bounding box and the ith detected bounding box, and \(A_i^u\)  is the area of union between the \textit{ith} ground-truth bounding box and the \textit{ith} detected bounding box. If the \(IoU_i\) is above the given threshold for a candidate window, the window is selected for the next stage. The three-stage threshold values used for selecting the candidate windows were 0.6, 0.7 and 0.9 respectively. The face thumbnails obtained from this framework have a size of 160*160 pixels. These thumbnails are provided to the face recognition framework as input. 
\subsection*{Face Recognition}
FaceNet is a deep CNN model that extracts facial features in terms of 128-D Euclidean (L2) embeddings using a triplet-based loss function \cite{RNdistance}. The input to FaceNet model is the set of aligned images obtained from MTCNN. The network is trained such that the squared L2 distances in the embedding space directly correspond to face similarity. These embedding vectors can then be used as feature vectors for a classification model. We used the FaceNet model pre-trained on a subset of the MS-celeb-1M dataset which includes about 10 million images of 100,000 celebrities \cite{RN589}, which had an accuracy of 0.99 on the Labeled Faces in the Wild (LFW) dataset \cite{RN819}. The pre-trained model was used to extract features from patient thumbnails and to calculate the corresponding L2 embeddings for our training images. These embeddings are then used to train a linear Support Vector Machine (SVM) for classification, on a fixed number of images (n=100, ~ 7 seconds of video) of a patient, along with the same number of negative examples of non-patients. The tolerance value for stopping criterion was set to 0.001.
\section*{Appendix B}
\subsection*{Posture Classification}
While our recorded patient frames contain examples of functional status activities such as walking, sitting in bed, or sitting in chair, these activities are interspersed in an imbalanced and sparse manner throughout the video clips. To remedy this problem, besides patient data, we additionally recorded 90 minutes of video containing scripted functional activities performed by nonpatients in the same ICU rooms. Out of the 150,621 video frames, 74,924 frames are scripted, and 75,697 frames are taken based on actual ICU patients’ videos.  
The initial size of each frame was 1680x1050, which was reduced to 368x654 to accommodate the memory. We used a multi-person pose estimation model \cite{RN493} to localize anatomical key-points of joints and limbs. Most algorithms are single-person estimators \cite{RN579,RN578,RN580}, such that they first detect each person and then estimate the location of joints and limbs. The single-person approach suffers from early commitment problem when multiple people are in close proximity; if an incorrect detection is made initially, there is no point of return as this approach tracks the initial detection. Due to the small size of hospital rooms and the presence of multiple people (patient, doctors, nurses, visitors), we used the multi-person approach \cite{RN493}. It also allows us to decouple the runtime complexity from the number of people for real-time implementations. The multi-person pose estimation was performed using the real-time multi-person 2D pose estimation with part affinity fields. The part affinity fields are 2D vector fields that contain information about the location and direction of limbs with respect to body joints. Our pose detection model consists of two branches of a sequential prediction process, where one branch detects the locations of joints, and the other branch detects the association of those body joints, as limbs. Both branches consist of Fully Convolutional Neural networks (FCN) \cite{RN718}. A convolutional network, consisting of first 10 layers of VGG-19 \cite{RN820}, is used to generate a set of feature maps \textbf{F}. These feature maps are used as input to each branch of the first stage of the model. The first branch outputs a set of detection confidence maps \(S^1=\rho^1(F)\) and the second branch outputs a set of part affinity fields \(L^1=\phi^1(F)\) where \(S^1\) and \(\phi^1\) are the two branches of CNNs at the first stage. In the following stage, the outputs from the branches in the previous stage and the original image features \textbf{F} are combined and provided as inputs to the two branches of the next stage, for further refinement. The confidence maps and part affinity fields for the subsequent stages are calculated as in equation (3) and equation (4), respectively \cite{RN493}.
\begin{equation}
S^t=\rho^t(F, S^{t-1},L^{t-1}),\forall t\geq2
\end{equation}
\begin{equation}
L^t=\phi^t(F,S^{t-1},L^{t-1}),\forall t\geq2
\end{equation}
This process is followed for the \textit{t} stages of the network. We have used three stages of the network in our model.

This model has been pre-trained on the MPII Human Pose dataset for 144,000 iterations. It contains over 40K activities with annotated body joints \cite{RN604}. The final model provided a state-of-the-art mean average precision of 0.79 on MPII dataset. We used the lengths of body limbs and their relative angles as features for the classification model. We used estimated poses to detect the four functional activities. We got the best results with K-Nearest Neighbors for classification, with Minkowski distance metric and value of K equal to one.

During the poselet detection step, sometimes a few anatomical key-points were not detected. This led to the problem of missing values for some features in the data that were provided to the classification model. Most algorithms are not immune to missing values. Several methods can be used to impute missing values, including mean, median, mode, or amputation via k-nearest neighbors (k-NN) \cite{RN821}. The \textbf{K} nearest neighbors are found based on the distance with the remaining features between the different samples. Each missing value of a feature was imputed by the weighted average of the same feature of the k nearest neighbors, with a K value of three. The resulting poselets were then used to train and test the classification algorithm on our dataset. We used 80\% of our data for training, and 20\% for training. The ICU training data included 74,924 frames from the scripted dataset and 75,697 frames from the actual ICU patients. Test data comprised only actual patient data. The hyper-parameters of the classification algorithms were fine-tuned using GridSearchCV with five-fold cross-validation. Pipeline of posture recognition model is shown in Figure \ref{fig:Figure_5b}. 

\begin{figure}
    \centering
    \begin{subfigure}[t]{\textwidth}
        \centering
        \includegraphics[width=\linewidth]{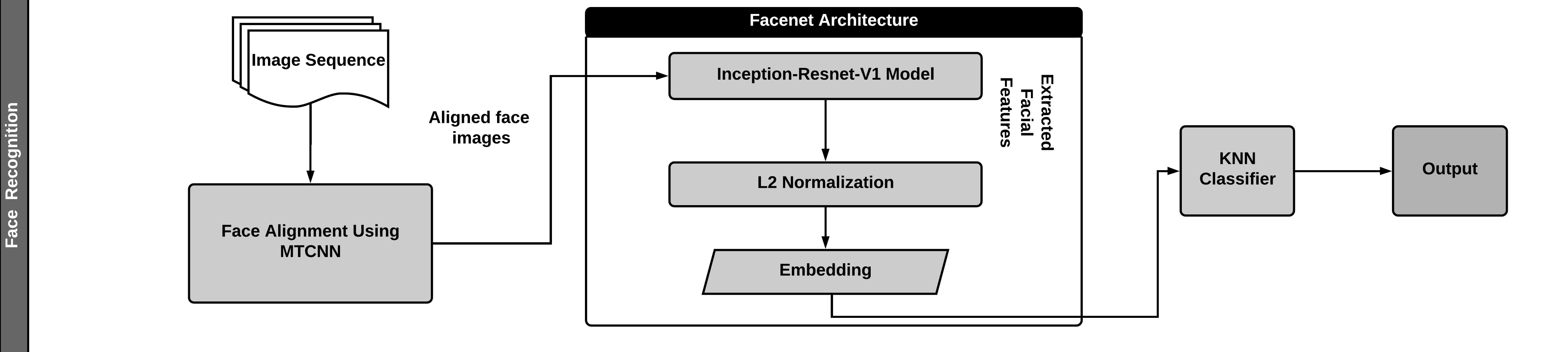} 
        \caption{} \label{fig:Figure_5a}
    \end{subfigure}
   \vspace{1cm}
    \begin{subfigure}[t]{\textwidth}
        \centering
        \includegraphics[width=\linewidth]{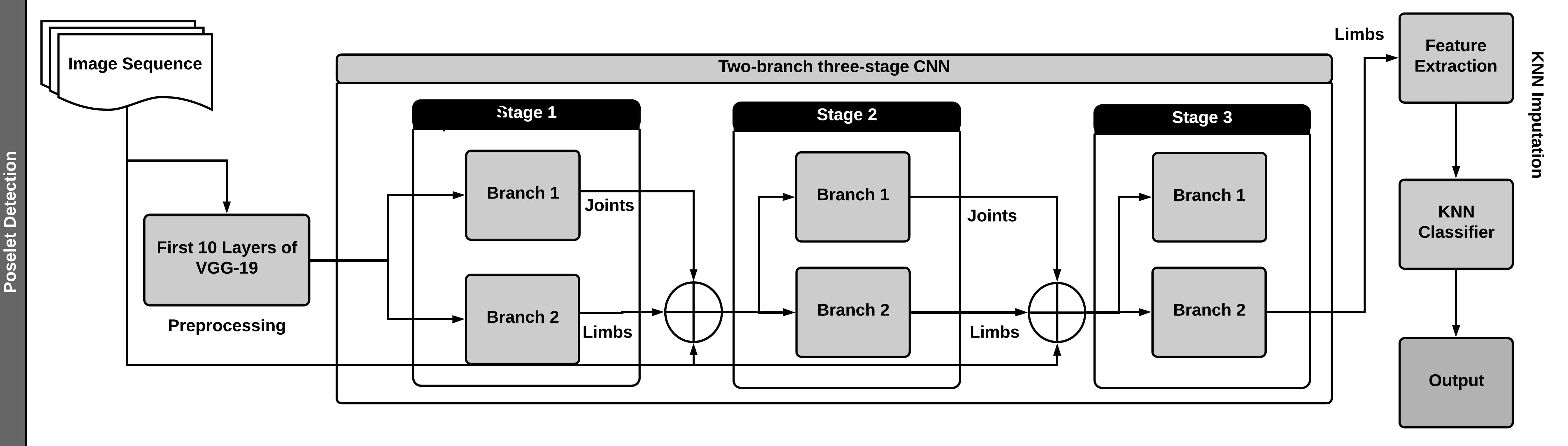} 
        \caption{} \label{fig:Figure_5b}
    \end{subfigure}

    \caption{a) Pipeline of the patient recognition system begins with alignment of faces in the image using MTCNN (Multi-stage CNN). The aligned images are then provided as input to the faceNet network which extracts features using a pre-trained Inception-Resnet-V1 model, then performs L2 normalization on them, and finally stores them as feature embeddings. These embeddings are used as input for the k-nearest neighbor (KNN) classifier to identify posture. b) Pipeline of the posture recognition system includes a two-branch three-stage CNN and a KNN classifier. At each stage of the CNN, Branch 1 predicts the confidence maps for the different body joints, and branch 2 predicts the Part Affinity Fields for the limbs. These predictions are combined at the end of a stage and refined over the subsequent stage. After stage 3, the part affinity fields of limbs are used to extract the lengths and angles of the body limbs. Any missing values are imputed using the KNN imputation, and a pre-trained KNN classifier is used to detect posture from the extracted features.}
\end{figure}

\begin{figure}[ht]
\centering
\setcounter{figure}{0}    
\makeatletter 
\renewcommand{\thefigure}{S\@arabic\c@figure}
\makeatother
\includegraphics[scale=0.17]{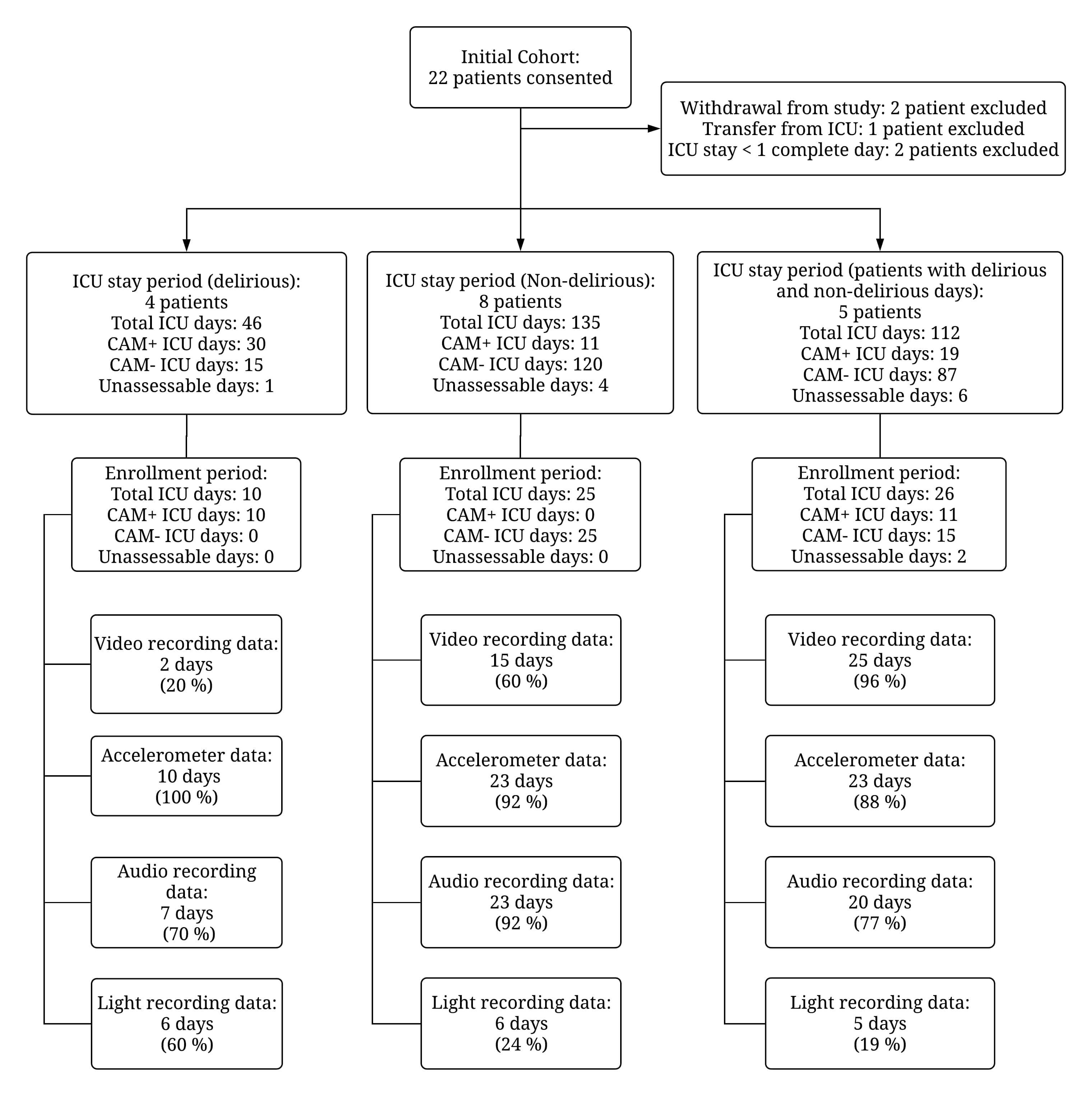}
\caption{Cohort recruitment diagram. Cohort recruitment diagram. Confusion Assessment Method for the Intensive Care Unit (CAM-ICU) was used to assess patients for delirium. If a patient had any positive CAM-ICU screening for a day, that day was identified as CAM+ (delirious). Days with no positive CAM-ICU screening, but which had negative CAM-ICU screening were identified as CAM- (Non-delirious). Days that the patient could not be assessed because of Richmond Agitation-Sedation Scale score of less than -3 were identified as unassessable days. Patients were divided into three groups: delirious patients (patients who were delirious through their enrollment period), non-delirious patients (patients who were not delirious through their enrollment period), and patients who had both delirious and non-delirious days. CAM: confusion Assessment Method, ICU: Intensive Care Unit.}
\label{fig:Figure_1_Supplement}
\end{figure}

\begin{figure}[ht]
\centering
\makeatletter 
\renewcommand{\thefigure}{S\@arabic\c@figure}
\makeatother
\includegraphics[scale=0.4]{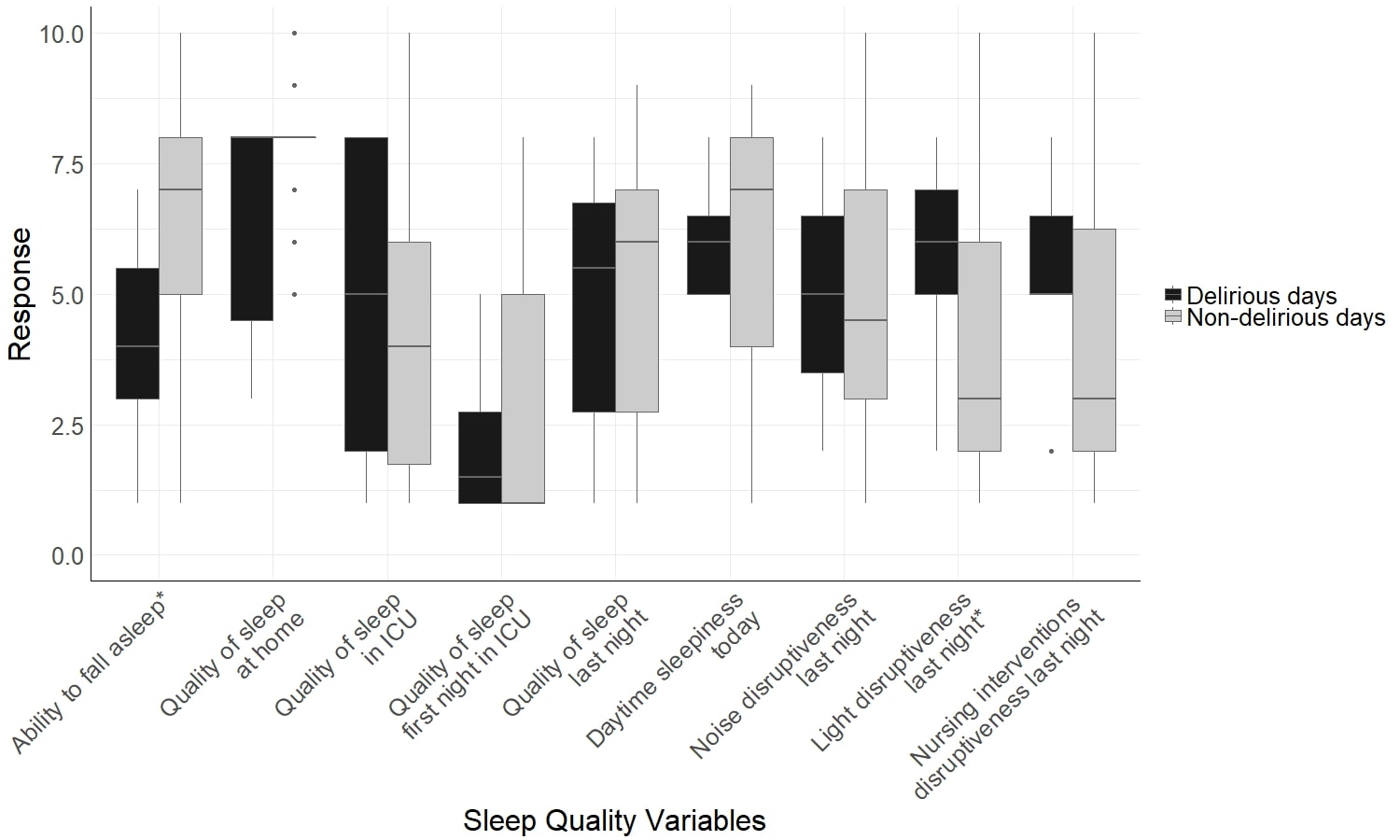}
\caption{Sleep quality outcomes, patient self-reports using Freedman Sleep Questionnaire. The parameters range from 1 to 10, with 1 being poor and 10 being excellent for the first five criteria. For overall daytime sleepiness, 1 is unable to stay awake, 10 is fully alert and awake. For environment and nursing interventions disruptiveness variables, 1 is no disruption, 10 is significant disruption. *: p-value less than 0.05. Number of delirium nights: 9. Number of non-delirium nights: 43.}
\label{fig:Figure_2__Supplement}
\end{figure}

\begin{figure}[ht]
\centering
\makeatletter 
\renewcommand{\thefigure}{S\@arabic\c@figure}
\makeatother
\includegraphics[scale=0.3]{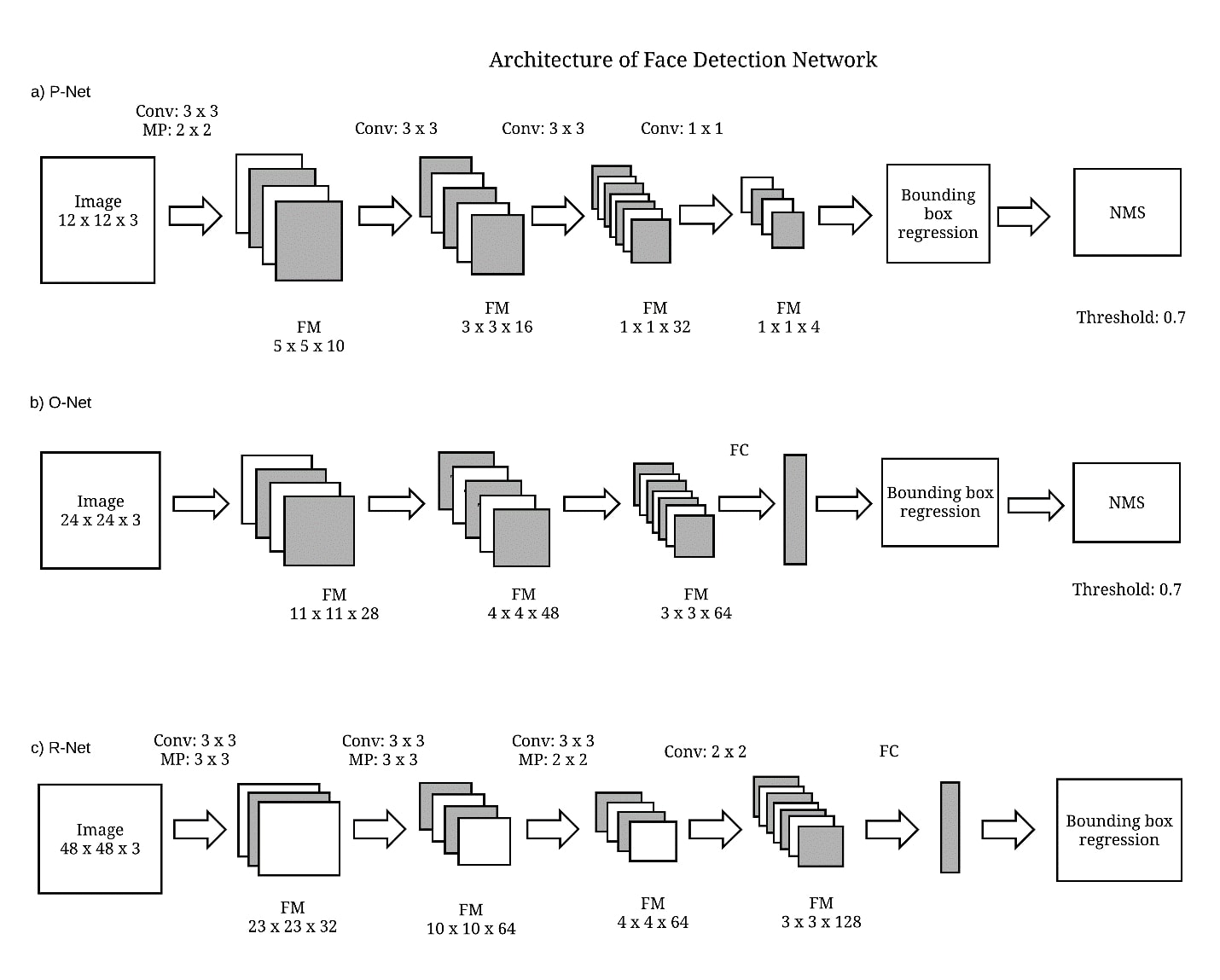}
\caption{Architecture Face detection network. a) Proposal network (P-Net) produces candidate windows possibly containing faces, b) Refine network (R-Net) rejects candidate windows not containing faces and performs bounding box regresison, c) Output network (O-Net) produces the final bounding box. Conv: Convolutional, MP: Max pooling, FC:  Fully Connected layer, FM:  Feature Maps and NMS:  Non-Maximum Suppression. The numbers denote the kernel size in Conv and MP layers. The numbers for FM denote the height, width and depth of the FM. The step-size for each Conv layer is one and for each MP layer is two.}
\label{fig:Figure_3__Supplement}
\end{figure}

\begin{table}[ht]
\centering
\begin{tabular}{|l|l|l|l|l|}
\hline
 & All Participants & Non-delirious (N=8) & Delirious (N=4) & p value\\
\hline
{Age, median (IQR)} & 69.0 (54.0, 73.0) & 62.5 (37.7, 73.0) &72.50 (64.5, 74.5) & 0.20 \\
\hline
{Female, number (\%)} & 4 (23.5) & 2 (25) & 1 (25) & 1\\
\hline
{Race, number (\%)} & & & & 1\\
\hline
{BMI, median (IQR)} & 26.2 (20.0, 29.7) & 28.4 (25.9, 30.7) & 29.5 (26.9, 32.7) & 0.81\\
\hline
{APACHE II, median (IQR)} & 26 (19, 39) & 20.5 (17.7, 27.7) & 35.5 (24.7, 45.5) & 0.12\\
\hline
{SOFA score at admission, median (IQR)} & 4 (2, 6) & 5 (2.5, 7.25) & 3.5 (2, 5.5) & 0.73\\
\hline
{Number of comorbidities, median (IQR)} & 1 (0, 4) & 3.5 (1.75, 7.25) & 0 (0, 0) & 0.02\\
\hline
{ICU LOS, median (IQR)} & 1 (0, 4) & 3.5 (1.75, 7.25) & 0 (0, 0) & 0.02\\
\hline
{ICU-free days, median (IQR)} & 6.9 (5.4, 10.2) & 7.6 (4.6, 18.2) & 6.2 (5.3, 7.9) & 0.68\\
\hline
{Hospital LOS, median (IQR)} & 16.5 (10.5, 22.0) & 27.0 (17.2, 44.2) & 17.5 (16.5, 19.2) & 0.28\\
\hline
{Death, number (\%)} & 3 (17.6) & 2 (25) & 0 (0) & 0.78\\
\hline
\end{tabular}
\caption{\label{tab:Table1}Cohort characteristics.}
\end{table}

\begin{table}[ht]
\centering
\begin{tabular}{|l|p{4cm}|p{4cm}|l|}
\hline
Variable, median (IQR) & Non-delirious patient days \newline(N=15) & Delirious patient days \newline(N=6) & p value \\
\hline
Mean activity count 24-hour & 53.9 (19.5, 161.6) & 291.3 (222.3, 426.2) & 0.02 \\
\hline
Standard deviation of activity count 24-hour & 199.9 (116.3, 456.3) & 540.8 (489.8, 759.4) & 0.04 \\
\hline
Mean activity count daytime & 69.7 (16.7, 198.1) & 333.1 (166.1, 375.4) & 0.04 \\
\hline
Standard deviation of activity count daytime & 246.8 (99.3, 472.5) & 564.0 (340.5, 683.0) & 0.07 \\
\hline
Mean of activity count nighttime & 46.4 (22.8, 94.7) & 321.5 (199.1, 489.4) & 0.01 \\
\hline
Standard deviation of activity count nighttime & 192.5 (130.5, 313.3) & 566.6 (451.1, 864.2) & 0.01 \\
\hline
M10 & 60137.3 (15029.5, 176498.3) & 268823.7 (253922.4, 413815.2) & 0.02 \\
\hline
Time of M10 (hour) & 6 (3, 9) & 8 (2, 13.2) & 0.58 \\
\hline
L5 & 3916.7 (1195.7, 10236.2) & 244016 (2621.9, 52125.3) & 0.38 \\
\hline
Time of L5 (hour) & 5 (3.5, 16.5) & 11 (1.7, 18.7) & 0.66 \\
\hline
Relative amplitude & 0.9 (0.7, 0.9) & 0.9 (0.8, 1) & 0.68 \\
\hline
RMSSD & 223.4 (137.3, 469.7) & 513.0 (475.7, 682.3) & 0.05 \\
\hline
RMSSD/SD & 1.1 (1.0, 1.2) & 0.9 (0.9, 1) & 0.07 \\
\hline
Number of immobile minutes daytime & 564 (416, 654) & 364.5 (236.2, 432) & 0.03\\
\hline
Number of immobile minutes nighttime & 602 (580, 650) & 359 (321.5, 483.5) & <0.01\\
\hline
\end{tabular}
\caption{\label{tab:Table2}Movement features for the wrist, compared between the delirious and non-delirious groups.}
\end{table}

\begin{table}[ht]
\renewcommand\thetable{1}
\renewcommand{\thetable}{S\arabic{table}}   
\setcounter{table}{0}
\centering
\begin{tabular}{|l|l|}
\hline
Facial Expression & AUs  \\
\hline
Happiness & 6+12 \\
\hline
Sadness & 1+4+15 \\
\hline
Surprise & 1+2+5+26 \\
\hline
Fear & 1+2+4+5+7+20+26 \\
\hline
Anger & 4+5+7+23 \\
\hline
Disgust & 9+15+16 \\
\hline
Contempt & R12A+R14A \\
\hline
Pain & 4+6||7+9||10+43 \\
\hline
\end{tabular}
\caption{\label{tab:Table 2_s}Action Units (AUs) for each facial expression.}
\end{table}

\begin{table}[ht]
\renewcommand{\thetable}{S\arabic{table}}   
\centering
\begin{tabular}{|l|l|l|}
\hline
Facial Action Unit Name & Facial Action Unit number & Binary/intensity coding \\
\hline
Inner brow raiser & AU1 & Intensity \\
\hline
Outer brow raiser & AU2 & Intensity \\
\hline
Brow lowerer & AU4 & Intensity/Binary \\
\hline
Upper lip raiser & AU5 & Intensity\\
\hline
Cheek raiser & AU6 & Intensity \\
\hline
Nose wrinkler & AU9 & Intensity \\
\hline
Lip corner puller & AU12 & Intensity/Binary \\
\hline
Dimpler &AU14 & Intensity \\
\hline
Lip corner depressor &AU15 & Intensity/Binary \\
\hline
Chin raiser &AU17 & Intensity \\
\hline
Lip stretcher &AU20 & Intensity \\
\hline
Lip tightener &AU23 & Binary \\
\hline
Lips part &AU25 & Intensity \\
\hline
Lip suck &AU28 & Binary \\
\hline
Blink &AU45 & Binary \\
\hline
\end{tabular}
\caption{\label{tab:Table 1_s}Action Units (AUs) detected using the OpenFace toolbox.}
\end{table}

\begin{table}[ht]
\renewcommand{\thetable}{S\arabic{table}}   
\centering
\begin{tabular}{|l|l|l|l|l|}
\hline
                            & & \multicolumn{3}{c}{Predicted label} \\
                            \hline
                            & & Lying & Sitting on chair & Standing \\
                            \hline
\multirow{3}{*}{True label} & Lying & 94.45 & 0.79 & 4.76 \\
                            & Sitting on chair & 1.73  & 92.89 & 5.38 \\
                            & Standing & 4.23  & 11.97 & 83.80\\   
                            \hline
                            
\end{tabular}
\caption{\label{tab:Table 3_s}Confusion matrix showing the model performance for the four postures -lying, sitting on bed, sitting on chair, and standing- using K-Nearest Neighbor model.}
\end{table}

\begin{table}[ht]
\renewcommand{\thetable}{S\arabic{table}}   
\centering
\begin{tabular}{|l|p{4cm}|p{4cm}|l|}
\hline
Variable, median (IQR) & Non-delirious patient days \newline(N=15) & Delirious patient days \newline(N=3) & p value \\
\hline
Mean activity count 24-hour & 25.6 (13.6, 125.9) & 4.8 (3.2, 15.2) & 0.10 \\
\hline
Standard deviation of activity count 24-hour & 106.6 (84.5, 346.1) & 81.6 (52, 95) & 0.20 \\
\hline
Mean activity count daytime & 33.9 (11.8, 126.9) & 6.7 (4.5, 11.9) & 0.08 \\
\hline
Standard deviation of activity count daytime & 139.1 (71.6, 370.1) & 54.3 (43, 84.8) & 0.08 \\
\hline
Mean of activity count nighttime & 21.8 (8.9, 66.1) & 0 (0, 19.8) & 0.15 \\
\hline
Standard deviation of activity count nighttime & 103.4 (58.1, 296.2) & 0 (0 , 70.9) & 0.12 \\
\hline
M10 & 30081.5 (13732.6, 147613.8) & 6841.1 (4636.5, 13373.4) & 0.06 \\
\hline
Time of M10 (hour) & 317 (162, 548) & 413 (241.5, 545.5) & 0.82 \\
\hline
L5 & 927.5 (593.4, 2789.8) & 0 (0, 252.7) & 0.04 \\
\hline
Time of L5 (hour) & 7 (8, 18) & 1 (1,4) & 0.15 \\
\hline
Relative amplitude & 0.9 (0.9, 1) & 1 (0.97, 1) & 0.06 \\
\hline
RMSSD & 117.6 (103.1, 360.7) & 85.6 (56.7, 102.8) & 0.20 \\
\hline
RMSSD/SD & 1.1 (1, 1.2) & 1.1 (1.1, 1.2) & 0.57 \\
\hline
Number of immobile minutes daytime & 589 (498.5, 670.5) & 683 (636.5, 697) & 0.16\\
\hline
Number of immobile minutes nighttime & 632 (601.5, 673) & 720 (605, 720) & 0.29\\
\hline
\end{tabular}
\caption{\label{tab:Table 4_s}Movement features for the arm, comparing between the delirious and non-delirious groups.}
\end{table}

\begin{table}[ht]
\renewcommand{\thetable}{S\arabic{table}}   
\centering
\begin{tabular}{|l|p{4cm}|p{4cm}|l|}
\hline
Variable, median (IQR) & Non-delirious patient days \newline(N=15) & Delirious patient days \newline(N=6) & p value \\
\hline
Mean activity count 24-hour & 8 (7.1, 27.1) & 16.6 (8.5, 53.0) & 0.46 \\
\hline
Standard deviation of activity count 24-hour & 61.4 (52.3, 91.2) & 57.7 (46.2, 138.8) & 0.91 \\
\hline
Mean activity count daytime & 8.9 (6.6, 28.3) & 17.3 (8.7, 57.9) & 0.51 \\
\hline
Standard deviation of activity count daytime & 61.7 (57.2, 98.7) & 60.4 (53.9, 130.5) & 0.85 \\
\hline
Mean of activity count nighttime & 9.8 (5.3, 22.4) & 22.4 (8.1, 46.8) & 0.51 \\
\hline
Standard deviation of activity count nighttime & 64.6 (42.2, 80.2) & 61.8 (32.0, 146.4) & 0.91 \\
\hline
M10 & 8094.1 (6817.2, 27183.4) & 18702.3 (10107.9, 44256.8) & 0.23 \\
\hline
Time of M10 (hour) & 6 (2.5, 8.5) & 9 (3.2, 13.2) & 0.56 \\
\hline
L5 & 544.4 (287.6, 2067.1) & 1226.5 (555.5, 8623.2) & 0.29 \\
\hline
Time of L5 (hour) & 13 (1.5, 16) & 10 (5, 18) & 0.69 \\
\hline
Relative amplitude & 0.9 (0.9, 0.9) & 0.8 (0.7, 0.9) & 0.15 \\
\hline
RMSSD & 75.6 (66.6, 109.9) & 60.4 (50.1, 162.4) & 0.56 \\
\hline
RMSSD/SD & 1.2 (1.1, 1.3) & 1.1 (1.1, 1.2) & 0.39 \\
\hline
Number of immobile minutes daytime & 650 (529.5, 686) & 601.0 (429.5, 648.7) & 0.35\\
\hline
Number of immobile minutes nighttime & 673 (544.5, 690.5) & 542 (523.5, 644.5) & 0.20\\
\hline
\end{tabular}
\caption{\label{tab:Table 5_s}Movement features for the ankle, comparing between the delirious and non-delirious groups.}
\end{table}

\begin{table}[ht]
\renewcommand{\thetable}{S\arabic{table}}   
\centering
\begin{tabular}{|l|l|p{5cm}|}
\hline
Sources at 1m & Sound Pressure & Sound Pressure Level \newline(reference sound pressure = 0 dB) \\
\hline
Threshold of pain & 20 Pa & 120 dB \\
\hline
Pneumatic hammer & 2 Pa & 100 dB \\
\hline
Street traffic & 0.2 Pa & 80 dB \\
\hline
Talking & 0.02 Pa & 60 dB \\
\hline
Library & 0.002 Pa & 40 dB \\
\hline
TV studio & 0.0002 Pa & 20 dB \\
\hline
Threshold of hearing & 0.00002 Pa & 0 dB \\
\hline

\end{tabular}
\caption{\label{tab:Table 6_s}Examples of sound pressure and sound pressure levels.}
\end{table}

\end{document}